\documentclass[final,1p,times,authoryear]{elsarticle}
\usepackage{float}
\usepackage[utf8]{inputenc}
\usepackage{amsmath,amsfonts,amssymb}
\usepackage{graphicx}
\usepackage{hyperref}
\usepackage{cleveref}
\usepackage{algorithm}
\usepackage{algorithmic}
\usepackage{booktabs}
\usepackage{threeparttablex}
\usepackage{tabularx}
\usepackage{siunitx}
\usepackage{orcidlink}
\usepackage{adjustbox}

\usepackage{xspace}
\newcommand{\COtwo}{CO\texorpdfstring{\textsubscript{2}}{2}\xspace}

\makeatletter
\def\ps@pprintTitle{%
  \let\@oddhead\@empty
  \let\@evenhead\@empty
  \def\@oddfoot{\reset@font\hfil\thepage\hfil}
  \let\@evenfoot\@oddfoot
}
\makeatother 

\begin{document}

\begin{frontmatter}

\title{Quantifying the Climate Risk of Generative AI: Region-Aware Carbon Accounting with G-TRACE and the AI Sustainability Pyramid}

\author[1]{Zahida Kausar}
\author[1]{Seemab Latif} 
\author[2]{Raja Khurram Shahzad\orcidlink {0000-0003-2806-9694}}
\author[1]{Mehwish Fatima\orcidlink {0000-0003-3424-2991}} 
\address[1]{School of Electrical Engineering and Computer Science (SEECS), National University of Sciences and Technology (NUST), Islamabad, Pakistan}
\address[2]{Department of Communication, Quality Management and Information Systems (KKI), Mid Sweden University, {\"O}stersund Campus, Sweden}
\begin{abstract}
Generative Artificial Intelligence (GenAI) represents a rapidly expanding digital infrastructure whose energy demand and associated \COtwo emissions are emerging as a new category of climate risk. This study introduces GenAI Transformative Carbon Estimator ({G-TRACE}), a cross-modal, region-aware framework that quantifies training- and inference-related emissions across modalities and deployment geographies. Using public activity traces, stratified annotation, de-duplication, and microscopic simulation, G-TRACE estimates energy use and carbon intensity per output type (text, image, video) and shows how decentralized inference can amplify small per-query energy costs into system-level impacts. Through the Ghibli-style image generation trend (2024–2025), we estimate 4,309 MWh of energy consumption and 2,068 metric tons \COtwo emissions, illustrating how viral participation inflates individual digital actions into tonne-scale consequences. Building on these findings, we propose the AI Sustainability Pyramid, a seven-level governance model linking carbon accounting metrics (L1–L7) with operational readiness, optimization, and stewardship. This framework translates quantitative emission metrics into actionable policy guidance for sustainable AI deployment. This study contributes to the quantitative assessment of emerging digital infrastructures as a novel category of climate risk, supporting adaptive governance for sustainable technology deployment.
By situating GenAI within climate-risk frameworks, the work advances data-driven methods for aligning technological innovation with global decarbonization and resilience objectives.
\end{abstract}
\end{frontmatter}
\section*{Keywords}
Generative AI; Carbon accounting; Climate risk; Sustainable digital infrastructures; Green AI; Energy-aware governance; Regional emissions; Policy frameworks.
\section{Introduction}
Generative Artificial Intelligence (GenAI) denotes transformer-based systems that process and generate multiple modalities (text, images, audio, video, and music) with human-like quality and creativity. GenAI has evolved rapidly from prototypes into widely deployed production tools. Across design, marketing, journalism, and entertainment, organizations adopt GenAI at scale; marketing and sales adoption more than doubles from 2023 to 2024~\citep{mckinsey2024state,mckinsey2023economic,karvonen2025marketing}. Prompt-based interfaces reduce technical barriers and allow non-experts to control complex models using natural language prompts.

Pivotal to this evolution are Large Language Models (LLMs) such as GPT-4~\citep{achiam2023gpt}, LLaMA~\citep{touvron2023llama}, Claude~\citep{bai2022constitutional}, Gemini~\citep{comanici2025gemini}, $DS$~\citep{liu2024deepseek}, and Grok~\citep{xai2024grok,de2025grok}. These systems interpret textual and voice prompts to generate content aligned with user intent. Furthermore, Vision Language Models (VLM), diffusion models DALL\mbox{·}E~\citep{marcus2022very}, Stable Diffusion~\citep{rombach2022high}, and Grok-Imagine~\citep{xai2024grok,de2025grok} are transforming visual content creation. Meanwhile, other emerging multimodal stacks are extending their capabilities toward text-to-video and voice-to-video generation. This expansion elevates computational and environmental concerns. ~\citet{vanderbauwhede2024estimating} estimates that a single ChatGPT query consumes roughly $60\times$ the energy of a standard web search; as usage scales, associated \COtwo emissions grow correspondingly~\citep{jiang2024revisit,arxiv2025optimizing,mit2025sustainability,gao2025environmental}.

\begin{figure}[t]
    \centering
    \includegraphics[width=\textwidth,keepaspectratio]{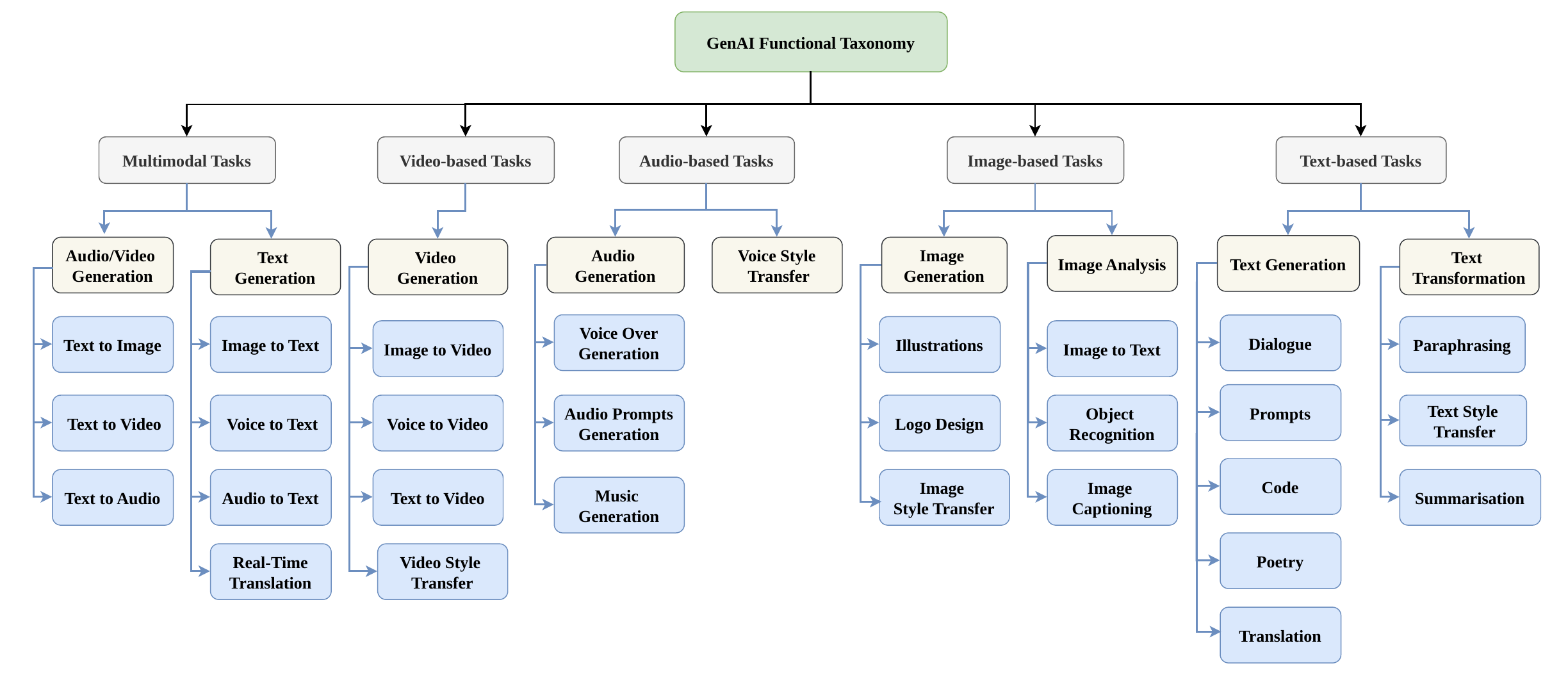}
    \caption{Overview of generative tasks and modalities (Multimodal, Video, Audio, Image, and Text).}
    \label{fig1}
\end{figure}

Figure~\ref{fig1} presents GenAI capabilities across five domains: (i) Multimodal, (ii) Video, (iii) Audio, (iv) Image, and (v) Text. Each domain exhibits distinct computational drivers (token length for text, resolution and diffusion steps for image, frames-per-second and duration for video) and therefore distinct energy and per-inference \COtwo profiles. Energy consumption of GenAI models scales with architecture and parameter count~\citep{tripp2024measuring,delrey2023energy,aquino2025towards}. We use this task–modality mapping as an analytical structure for cross-modal comparisons and for isolating energy-intensive operations that demand targeted mitigation.

To minimize emissions, leading technology companies announce ambitious sustainability goals.\footnote{Examples include: Google's \textit{24/7 Carbon-Free Energy by 2030} initiative (2020), Microsoft's \textit{Carbon Negative by 2030} commitment (2020), Meta's \textit{Sustainability Report} (2024), and Amazon's \textit{Carbon Methodology Report} (2023).} Yet GenAI workflows (iterative prompting, fine-tuning, and large-scale batch generation) strain these sustainability commitments. After deployment of these models, inference often dominates their life-cycle 
emissions~\citep{luccioni2023power,dodge2022measuring}. Approximately, one year of inference can emit $25\times$ the \COtwo of GPT-3 training due to billions of daily interactions~\citep{wu2023reducing}. These insights motivate to develop inference-efficient algorithms, architectures, and carbon-aware scheduling~\citep{strubell2020energy,luccioni2024impact}. The concept of  Green AI evolves with this motivation to develop optimized and energy-efficient AI workflows. Its agenda advances pruning, quantization, distillation, sparse Mixture-of-Experts~\citep{lin2024moe,han2015deep}, and hardware/software co-design.

Despite these efforts, critical gaps persist. Many metrics rely on proxy indicators like GPU-usage hours, and FLOPs (Floating-point Operations per second) for quantifying \COtwo emissions.  Other energy requirements like upstream manufacturing and downstream storage or cooling are usually ignored. Cultural accelerants (fandoms, remix culture, meme dynamics) create unanticipated energy demand spikes for GenAI sytems. To expose these dynamics, we analyze a viral, culturally anchored workload: Ghibli-style image generation and quantify its energy use and emissions across platforms and regions.\footnote{Ghibli-style: Denotes user-generated images that mimic the visual aesthetics of Studio Ghibli; the trend spreads globally across social platforms.}

\begin{figure}[t]
    \centering
    \includegraphics[width=\textwidth,keepaspectratio]{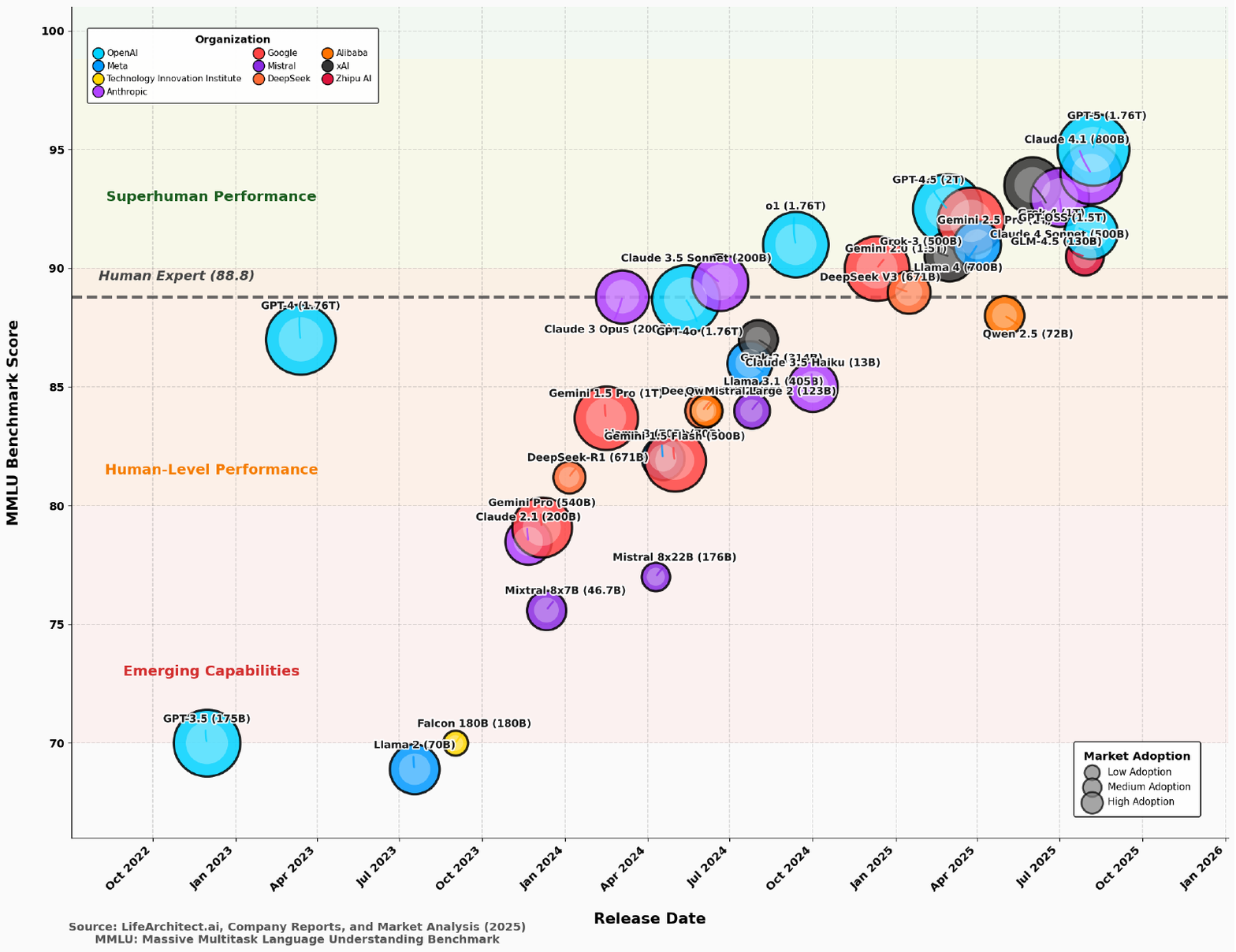}
    \caption{GenAI Model Performance Evolution: MMLU Performance Benchmarks vs. Model Release Timeline (2022–2025) with Market Adoption Indicators. Models are categorized into three performance tiers based on MMLU scores: Emerging, Human-Level, and Superhuman. This figure shows the accelerating advancement of GenAI models toward superhuman performance within a three-year development cycle.}
    \label{fig2}
\end{figure}

Figure~\ref{fig2} presents widely adopted models by performance and release date. Scaling along this curve requires substantial compute and energy resources. For example, training a single language model emits an estimated 626,000 pounds of \COtwo, comparable to the lifetime emissions of five gasoline-powered American cars~\citep{strubell2020energy}. Newer systems likely exceed this baseline estimation due to exponentially larger model sizes and increased training dataset volumes~\citep{bommasani2021opportunities}. With atmospheric \COtwo at record highs in 2023~\citep{shukla2022climate}, sustainable GenAI design shifts from aspiration to requirement. Against this backdrop, we select the $Ghix$ trend to illustrate how distributed, user-driven adoption produces measurable emissions at scale.
\subsection*{Contributions} 
This work delivers a \textit{measurement-to-action} pipeline that couples region-aware carbon accounting with empirical demand signals and operational guidance.
\begin{enumerate}
    \item We present \textit{G-TRACE} (GenAI TRAnsformative Carbon Estimator), a region-aware carbon accounting framework that quantifies training- and inference-related emissions across modalities, normalizes by output type (tokens, images, video duration), and combines energy-use estimates with regional grid factors to calculate location-specific carbon footprints.
    \item We calibrate G-TRACE using a dual approach: (i) macroscopic platform analytics that capture $Ghix$ activity signals as noisy observables, corrected via stratified annotation and de-duplication, and (ii) microscopic simulations that estimate per-output energy $e_{\mathrm{img}}$ across device classes, frameworks, and resolutions.
   \item We apply G-TRACE to the $Ghix$ trend (2024--2025) and estimate a range of electricity use and \COtwo emissions under  explicit assumptions, showing how viral participation can convert small per-query costs into ton-scale impacts.
    \item We translate these measurements into operational policy via an AI Sustainability Pyramid (readiness levels L1–L7) that sets concrete gates for measurement and monitoring, optimization, portfolio governance, ecosystem standards, and net-positive stewardship.
\end{enumerate}

\subsection*{Paper Structure}
Section~\ref{sec:related} reviews technical, social, and environmental literature. Section~\ref{sec:energy} examines the energy use and carbon impacts of GenAI. Section~\ref{sec:framework} presents G-TRACE and its region-aware accounting boundary. Section~\ref{sec:case} applies G-TRACE to $Ghix$ and analyzes sensitivity. Section~\ref{sec:validation} reports macro-micro validation and parameter calibration. Section~\ref{sec:results} presents the core results and findings. Section~\ref{sec:pyramid} operationalizes findings via the AI Sustainability Pyramid. Section~\ref{sec:discussion} discusses implications, and Section~\ref{sec:conclusion} concludes.

\section{Literature Review}
\label{sec:related}
This section examines GenAI through four interconnected perspectives: technological evolution, socio-cultural transformation, environmental costs, and mitigation strategies under Green AI. We trace the evolution of multimodal capabilities and highlight key breakthroughs across text~\citep{achiam2023gpt,touvron2023llama,comanici2025gemini}, image~\citep{marcus2022very,rombach2022high}, audio~\citep{borsos2023audiolm,agostinelli2023musiclm,kreuk2023audiogen}, and video generation~\citep{ho2022video} that transform creative and professional workflows~\citep{vives2023artificial,mckinsey2024state,karvonen2025marketing}. Next, we analyze socio-ethical implications of widespread adoption, focusing on access disparities, algorithmic bias, and creative labor displacement~\citep{siau2020artificial,torres2025language,bai2022constitutional,mckinsey2023economic,sonni2024digital}. We then analyze environmental costs of training and inference~\citep{strubell2020energy,henderson2020systematic,patterson2021carbon,luccioni2023estimating,luccioni2023power,samsi2023words,wu2023reducing} and finally we review emerging Green AI governance frameworks~\citep{schwartz2020green,verdecchia2023systematic,kaack2022aligning,bommasani2021opportunities,hacker2023ai,alder2024ai}. Together, these perspectives reveal how technical innovation, cultural dynamics, and sustainability imperatives shape GenAI's trajectory.

\subsection{Evolution of GenAI}
The GenAI landscape evolves rapidly from single-modality autoregressive models (designed for token-by-token text generation) to advanced multimodal systems, enabled by exponential increases in computational capacity driven by specialized GPU architectures.\footnote{Yoffie, D.B. and Von Bargen, S. (2024). \textit{Nvidia, Inc. in 2024 and the Future of AI}. Harvard Business School Case 725-360, September 2024.} Recent models demonstrate agentic behavior and cross-domain generation. This progression runs from early transformer architectures ($EG$, GPT-3) to instruction-tuned, multimodal Large Language Models (LLMs) including $GPTfx$~\citep{achiam2023gpt}, $LMAx$~\citep{touvron2023llama}, $Cl-4$~\citep{bai2022constitutional}, $Gemx$~\citep{comanici2025gemini}, and $DS$~\citep{liu2024deepseek}. These systems perform contextual reasoning, maintain long-horizon semantic coherence, and operate as the intelligence layer in digital assistants, programming copilots, and autonomous agents.

In the visual domain, diffusion-based Vision–Language Models (VLMs) such as $DALx$~\citep{marcus2022very} and $SDx$~\citep{rombach2022high} enable high-fidelity, style-adaptive image synthesis. They produce highly stylized outputs, including Ghibli-style artwork, fueling remix cultures that blur the boundary between tool (GenAI model) and artist (human creator). Video synthesis also advances significantly, models such as $Sox$~\citep{fadiya2025sora} enable prompt-driven cinematic video with temporal coherence and strong perceptual quality~\citep{ho2022video}.

While these capabilities deliver unprecedented creative efficiency and broaden access to professional-quality content~\citep{mckinsey2023economic,vives2023artificial}, they simultaneously rely on resource-intensive backend powered by high-performance GPU clusters, large memory footprints, and massive datasets that draw substantial energy~\citep{strubell2020energy,henderson2020systematic,patterson2021carbon,luccioni2023estimating}.
 As systems like $Sox$ reach global adoption, cumulative environmental impact scales with engagement~\citep{devries2023growing,vanderbauwhede2024estimating}. This disconnect between effortless experience and intensive computation creates a sustainability challenge that demands coordinated responses across research, development, and policy~\citep{kaack2022aligning,alder2024ai}. We next consider how these technical shifts reverberate socially and culturally.

\subsection{Socio-Cultural Impact of GenAI}
GenAI reconfigures creative production by removing traditional technical barriers for new entrants~\citep{mckinsey2024state,vives2023artificial,karvonen2025marketing}. Modern systems enable amateurs to produce professional multimedia without extensive formal training; industry surveys report 65–70\% adoption among designers in high-income economies~\citep{mckinsey2023economic}. Viral phenomena such as $Ghix$ showcase user-led remixing of established artistic styles through AI-generated images~\citep{torres2025language}, catalyzing participatory sharing cultures across social platforms.

However, this accessibility exposes stark geographic and economic divides. Creators in the Global South face inadequate infrastructure, high subscription costs, and limited access to high-performance hardware~\citep{siau2020artificial}. These divides amplify risks of cultural homogenization, intellectual-property infringement, and labor displacement~\citep{torres2025language,bai2022constitutional,mckinsey2023economic,sonni2024digital}. Western-centric training corpora bias outputs toward dominant aesthetics, while automated workflows displace tasks previously done by human creators. Competitive dynamics then privilege creators with premium AI access, intensifying inequality in global digital marketplaces~\citep{mckinsey2023economic,siau2020artificial}. These socio-cultural dynamics intersect with environmental outcomes, since access and infrastructure also shape where and how compute runs.

\subsection{Environmental Costs of GenAI Models}
\label{sec:envcosts}
Early research focused primarily on training emissions, but inference now dominates life-cycle \COtwo for many GenAI deployments~\citep{luccioni2023power,wu2023reducing}. Training large models ($EG$, GPT-3/4) emits roughly 200--626 metric tons of \COtwo~\citep{strubell2020energy,patterson2021carbon}. Inference scales across billions of daily interactions and can exceed training emissions over a model's operational lifetime~\citep{luccioni2023estimating,samsi2023words}. Regional grid composition drives further variance: hydro-dominant Norway operates at \SI{0.023}{kgCO2/kWh}, whereas coal-intensive India operates at \SI{0.819}{kgCO2/kWh}~\citep{chen2025comparative}, a $\sim$35.6$\times$ disparity for identical workloads. Deployment geography therefore critically shapes environmental impact~\citep{dodge2022measuring}.

These disparities arise from infrastructure inequities, with high-emission data centers concentrated in regions lacking renewable supply or advanced cooling~\citep{devries2023growing}. 
Carbon externalization occurs when users in low-emission regions trigger inference on hardware in high-emission zones. 
Transparency remains limited, as few vendors disclose region-specific emissions~\citep{henderson2020systematic}. 
Renewable initiatives such as India's Bhadla Solar Park (2.25~GW) offer pathways toward greener compute~\citep{somani2025rajasthan}, yet exponential GenAI adoption risks outpacing decarbonization~\citep{vanderbauwhede2024estimating}. 

\begin{table}[h]
\centering
\begin{adjustbox}{width=0.95\textwidth}
\begin{tabular}{lcccc}
\toprule
\textbf{Region} & \textbf{Grid EF (kg \COtwo/kWh)} & \textbf{Training (t \COtwo)} & \textbf{Inference (g/query)} & \textbf{Image (g/image)} \\
\midrule
\textbf{\textit{Norway (Best)}} & \textbf{0.023} & 110 & 0.42 & 3.8 \\
France & 0.052 & 248 & 0.95 & 8.6 \\
EU Avg. & 0.231 & 1,100 & 4.20 & 38.1 \\
USA & 0.379 & 1,810 & 6.90 & 62.5 \\
China & 0.577 & 2,750 & 10.50 & 95.2 \\
\textbf{\textit{India (Worst)}} & \textbf{0.819} & 3,910 & 14.90 & 135.1 \\
\bottomrule
\end{tabular}
\end{adjustbox}
\caption{Regional grid emission factors and GenAI \COtwo implications. 
\emph{Norway} represents the lowest-emission configuration (\textit{best case}), 
while \emph{India} shows the highest-emission configuration (\textit{worst case}). 
Large disparities motivate geographically adaptive deployment and carbon-aware scheduling.}
\label{tab:tb1}
\end{table}

Taken together, these findings motivate the need for end-to-end, region-aware accounting and governance under a Green AI agenda linking accurate measurement, equity-aware deployment, and operational levers for emission reduction~\citep{schwartz2020green,verdecchia2023systematic,kaack2022aligning}. The following section (\ref{sec:greenai}) builds upon this analysis by outlining architectural, algorithmic, and regulatory strategies that translate such measurements into concrete mitigation approaches.

\subsection{Green and Sustainable AI Solutions}
\label{sec:greenai}
As detailed in Section~\ref{sec:envcosts}, emissions quantification depends strongly on the regional carbon intensity of energy supply and the geographic distribution of computational workloads. Building upon that foundation, the Green AI movement emphasizes the joint optimization of accuracy and energy efficiency~\citep{schwartz2020green}. Several approaches curb emissions without sacrificing performance. Sparse Mixture-of-Experts (MoE) architectures activate only parameter subsets per query, lowering compute while preserving fidelity. Model compression techniques (pruning, quantization, knowledge distillation) reduce FLOPs and memory costs. Federated and edge inference shift computation closer to users, while dynamic voltage and frequency scaling (DVFS) reduces hardware power draw under variable load. Combined with \emph{carbon-aware scheduling and routing}, these design strategies align model deployment with local grid conditions and renewable availability.
Regulatory initiatives such as the EU AI Act have begun mandating energy transparency~\citep{act2024eu,lewis2025mapping}, but reporting remains inconsistent and often omits upstream processes (data collection, hardware production) and downstream user-driven inference. Achieving climate alignment requires full life-cycle coverage and decision-grade telemetry. Priority directions include: 
(i)~\emph{carbon traceability} through real-time, per-output reporting; 
(ii)~\emph{low-impact user experience design} that discourages unnecessary queries and high-resolution defaults; and 
(iii)~\emph{policy instruments} that incentivize low-emission deployments via pricing, procurement standards, and launch thresholds linked to intensity metrics.

\begin{table}[ht]
\centering
\begin{adjustbox}{width=0.95\textwidth}
\begin{tabular}{lllll}
\toprule
\textbf{Model} & \textbf{Params (B)} & \textbf{Training (MWh)} & \textbf{Training (t \COtwo)} & \textbf{Inference (Wh/query)} \\
\midrule
GPT-4 & 1,760 & 28,800 & 6,912 & 2.9 \\
LLaMA-2 (70B) & 70 & 1,270 & 539 & 1.8 \\
Claude 3 & 175$^*$ & 6,200$^*$ & 2,350$^*$ & 2.5$^*$ \\
Gemini Pro & 100$^*$ & 20,000$^*$ & 5,000$^*$ & 2.4$^*$ \\
Stable Diffusion & 1.0 & 300 & 72 & 15$^\dagger$ \\
DALL·E 2 & 12 & 1,000 & 240 & 10$^\dagger$ \\
PaLM (540B) & 540 & 6,400 & 1,500 & 4.5 \\
\bottomrule
\end{tabular}
\end{adjustbox}
\caption{Estimated training and inference emissions for major GenAI models. Values highlight modality-specific trade-offs: text-based LLMs dominate training costs, while diffusion image generators incur higher per-query inference energy. $^*$Approximate values; $^\dagger$per-image inference from diffusion studies.}
\label{tab:tb2}
\end{table}

As shown in Table~\ref{tab:tb2}, text-based LLMs dominate training costs, whereas diffusion image generators incur higher per-query inference energy. This highlights the need for sustainable AI design that integrates both technological efficiency and policy mechanisms to reduce emissions across the model lifecycle.

\subsection{Summary and Research Gaps}
The literature on GenAI reveals three overarching insights that collectively define its current trajectory. 
GenAI delivers unprecedented multimodal capabilities, spanning text, image, audio, and video generation, yet these advances depend on highly resource-intensive computational infrastructure. At the same time, socio-cultural impacts such as cultural homogenization, intellectual-property risks, and creative labor displacement remain unevenly distributed across regions, reinforcing existing technological inequalities. Environmental costs are also increasing across both training and inference phases, while mitigation efforts remain fragmented across algorithmic, architectural, and policy domains.

Despite growing attention to Green AI, current research still lacks a unified, region-aware framework that consistently quantifies carbon emissions across model modalities, deployment contexts, and geographic locations. 
Most studies rely on proxy indicators such as GPU-hours or FLOPs, treat carbon routing and grid intensity only superficially, and underestimate the amplification effect of viral user behavior that drives inference emissions to dominate the life cycle. Boundary definitions also vary widely, as many omit upstream embodied emissions or downstream storage and cooling, while limited vendor transparency continues to hinder comparability and governance.

These limitations highlight the need for a comprehensive and standardized accounting framework that bridges technical modeling, environmental measurement, and policy decision-making. To address this gap, we propose {G-TRACE}, a structured, cross-modal carbon accounting framework that integrates regional grid emission factors, workload distribution, and modality-specific energy profiles. We validate its parameters using macro and micro empirical evidence and apply it to a viral, culturally anchored workload ($Ghix$) to quantify inference-driven emissions at scale. Finally, we translate these measurements into actionable policy and organizational practices through the {AI Sustainability Pyramid}, which defines readiness stages for instrumentation, optimization, portfolio governance, and long-term stewardship.

\section{Energy and Carbon Impacts of GenAI}
\label{sec:energy}
The deployment of GenAI systems accelerates digital content creation, automation, and productivity~\citep{strubell2020energy,schwartz2020green}. However, this rapid expansion also intensifies environmental costs as energy demand rises for large-scale language, vision, and video models. This section examines the carbon footprint of GenAI across model families, usage patterns, and deployment infrastructures. We begin by characterizing cross-modal energy and \COtwo profiles of flagship models, followed by an analysis of scaling dynamics in training and inference. Finally we generalize a region-aware accounting framework that explains why identical workloads yield divergent emissions across geographies. 
These findings are later contextualized in the viral $Ghix${} case and linked to organizational governance via the AI Sustainability Pyramid.

\subsection{Cross-Modal Energy and \COtwo Profiles of Flagship Models}
Table~\ref{tab:tb2} summarizes representative training and inference footprints for leading GenAI models. 
For instance, OpenAI's GPT-4 (1.76T parameters) consumed approximately \SI{28800}{\mega\watt\hour} during training, resulting in about \SI{6912}{\tonne} of \COtwo, whereas LLaMA-2 (70B) required \SI{1270}{\mega\watt\hour} and emitted \SI{539}{\tonne}. Although per-query inference energy is much smaller, it scales rapidly in production: GPT-4 averages \SI{2.9}{\watt\hour\per\text{query}}, while Claude~3 and Gemini~Pro each require roughly \SIrange{2.4}{2.5}{\watt\hour\per\text{query}}~\citep{luccioni2023estimating,samsi2023words}. 
Image-generation systems such as $DALx$ typically require \SIrange{100}{200}{\mega\watt\hour} for training and produce \SIrange{50}{100}{\tonne} of \COtwo~\citep{marcus2022very}, with per-image emissions ranging from \SIrange{2}{4}{\gram\COtwo\per\text{image}} at default settings but increasing substantially with higher resolution and more diffusion steps.

Video synthesis models ($EG$, $Sox$) are substantially more energy-intensive, with inference costs approaching twenty times those of image synthesis due to temporal coherence and motion-frame modeling~\citep{gao2025environmental}. These values illustrate a consistent pattern: text-based LLMs dominate training-related emissions, while diffusion and video models exhibit higher per-query inference intensity. Together, they highlight that sustainability challenges in GenAI extend beyond architecture to encompass usage patterns and deployment scale.

\subsection{Scaling Trends Across Modalities}
Frontier LLMs such as GPT-4 continue to expand exponentially in both parameter count and computational cost~\citep{patterson2021carbon}. Estimates suggest that its training emissions are equivalent to driving a conventional gasoline car over 18 million miles. While smaller models such as LLaMA improve emissions-per-FLOP efficiency, sustained global usage has shifted the majority of life-cycle emissions toward inference, contributing nearly \SI{10}{\tera\watt\hour\per\mathrm{yr}} worldwide~\citep{luccioni2023estimating}. 

In contrast, visual and multimodal generators exhibit lower total training energy but greater per-query variability. Higher-resolution rendering, longer diffusion chains, and multimodal pipelines all amplify cumulative energy use. As adoption accelerates, these scaling dynamics reinforce the importance of system-level optimization, hardware efficiency, and adaptive workload scheduling.

\subsection{Region-Aware Carbon Accounting Framework}
To capture spatial variability in emissions, we extend the standard energy–emission formulation~\citep{shukla2022climate,lacoste2019quantifying,patterson2021carbon,souza2023casper}:
\begin{equation}
m_{\mathrm{CO_2}}\,[\si{kg}] = E\,[\si{kWh}] \times EF\,[\si{kgCO2\per kWh}],
\label{eq:region}
\end{equation}
and its regionalized version:
\begin{equation}
m_{\mathrm{CO_2}} = \sum_{k \in \mathcal{K}} \big(E_{\mathrm{train},k} + E_{\mathrm{infer},k}\big)\, EF_k,
\label{eq:region_weighted}
\end{equation}
where \(m_{\mathrm{CO_2}}\) denotes emitted carbon dioxide, \(E\) represents energy consumed during training and inference, and \(EF_k\) is the regional grid emission factor. 
This formulation explains the large spread observed in Table~\ref{tab:tb1}, where identical computational workloads produce markedly different emissions depending on siting, routing, and energy-source composition. 

Building upon this framework~\citep{somani2025rajasthan,henderson2020systematic,lacoste2019quantifying}, we integrate local grid factors with workload-specific energy data to enable region-sensitive estimation across deployment scenarios.These principles underpin the operational analysis in later sections, where region-aware allocation informs sustainable scheduling and governance through the AI Sustainability Pyramid.

\section{G-TRACE: A Cross-Modal, Region-Aware Carbon Accounting Framework for GenAI}\label{sec:framework}
This section introduces {GenAI TRAnsformative Carbon Estimator}, a cross-modal, region-aware framework that quantifies the environmental footprint of GenAI workloads across heterogeneous device ecosystems. Unlike compute-only accounting, we integrate social-trend analytics with device-level energy modeling and region-specific factors. G-TRACE quantifies both training and inference, normalizes results by modality, and outputs uncertainty-aware metrics.

\begin{figure*}[ht]
\centering
\includegraphics[width=0.8\columnwidth,keepaspectratio]{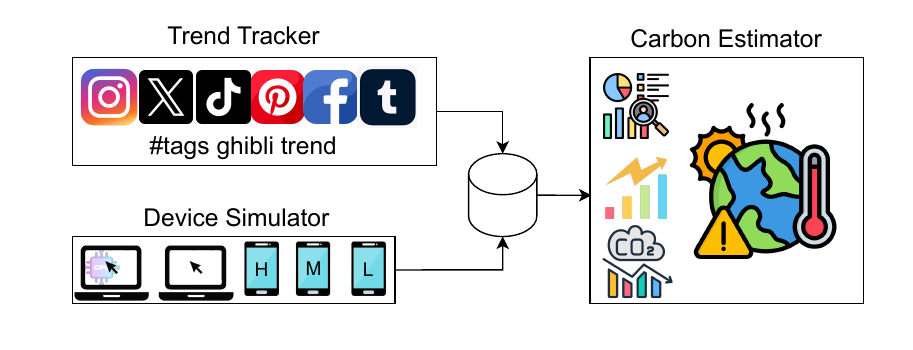}
\caption{G-TRACE framework architecture. Three stages, i.e., {Trend Tracker}, {Device Simulator}, and the region-aware {\COtwo Estimator}—convert social activity signals and workload metadata into energy and \COtwo estimates with uncertainty bounds. Device tiers span Laptop with and without GPUs and mobile classes (H/M/L). Region-specific grid factors enable fair, cross-modal comparisons and sensitivity analyses.}
\label{fig3}
\end{figure*}

G-TRACE operates through three integrated stages: (i) {Trend Tracker} for viral trend identification and data acquisition, (ii) {Device Simulator} for heterogeneous energy modeling, and (iii) {\COtwo Estimator} with uncertainty analysis. Figure~\ref{fig3} illustrates the pipeline: we first select a trend ($EG$, Ghibli-style hashtag counts), the {Trend Tracker} collects data across social media platforms, and the {Device Simulator} reproduces the workload in a controlled environment across laptop devices with(and without) GPUs and three mobile tiers—high-resource (H), medium-resource (M), and low-resource/edge (L). The region-weighted {\COtwo Estimator} then produces comparable cross-modal metrics and supports sensitivity studies. 

Below, we describe the G-TRACE pipeline end-to-end. While the framework centers on three integrated components, i.e., {Trend Tracker}, {Device Simulator}, and a region-aware {\COtwo Estimator}—we organize the exposition into data acquisition, energy modeling, region-aware estimation, and comparative assessment for clarity.

\subsection{Data Acquisition}
G-TRACE ingests three input strata aligned to the system boundary. First, \emph{model/workload metadata} capture model family and scale (parameter count), training schedule, and inference hyperparameters ($EG$, diffusion steps, resolution, sampler; context length for LLMs; frame rate and duration for video), together with device class ($EG$, A100, H100) and datacenter PUE. Second, \emph{activity signals} quantify the volume and mix of generations by modality (text, image, video) over a defined time window and platform distribution; for $Ghix$, hashtag-level post counts across platforms (Table~\ref{tab:tb3}) serve as the image activity proxy. Third, \emph{region signals} specify deployment geographies and their grid emission factors (kg \COtwo/kWh), with optional traffic weights for multi-region routing. These streams (Figure~\ref{fig3}, left) parameterize the energy modeling step, which converts workload characteristics and activity volumes into training and inference energy vectors.

\subsection{Energy Consumption Modeling}
G-TRACE models training and inference separately and then aggregates. When telemetry is unavailable, training energy follows
\begin{equation}
E_{\mathrm{train}} \;=\; \Big(\sum_{d \in \mathcal{D}} P_d \, U_d \Big)\, T \cdot \mathrm{PUE},
\label{eq:train}
\end{equation}
where $P_d$ denotes device power (kW), $U_d$ average utilization, $T$ training time (h), and $\mathrm{PUE}$ facility overhead. Literature baselines adjust for model scale and epochs~\citep{strubell2020energy,patterson2021carbon}. Per-output inference energy depends on modality and configuration,
\begin{equation}
e_{\mathrm{img}}(r,s), \qquad e_{\mathrm{text}}(L), \qquad e_{\mathrm{vid}}(r,f,t),
\label{eq:per_output}
\end{equation}
with spatial resolution $r$, diffusion steps $s$, token length $L$, frame rate $f$, and duration $t$. Total inference energy aggregates over modality counts $N_m$ as
\begin{equation}
E_{\mathrm{infer}} \;=\; \mathrm{PUE} \cdot \sum_{m \in \{\text{text,img,vid}\}} N_m \, e_m(\cdot).
\label{eq:infer}
\end{equation}
Sensitivity analysis perturbs $(\mathrm{PUE}, e_m)$ within empirically grounded ranges; optional FLOPs-to-Wh mappings by device efficiency calibrate $e_m(\cdot)$. The resulting energy vectors feed the region-aware estimation (Figure~\ref{fig3}, center).

where $k$ indexes regions and $EF_k$ is the grid emission factor (kg \COtwo/kWh). For multi-region routing with weights $w_k$ and $\sum_k w_k=1$, we set $E_{\bullet,k}=w_k \, E_{\bullet}$ for $\bullet \in \{\mathrm{train},\mathrm{infer}\}$. Region factors derive from official inventories ($EG$, IEA), distinguishing low-intensity grids ($\sim 0.023$) from coal-heavy grids ($\sim 0.8$). Figure~\ref{fig3} (right) then aggregates and normalizes outputs into cross-modal metrics and lifecycle profiles.

G-TRACE reports absolute and normalized intensity metrics ($EG$, g \COtwo per image; g \COtwo per \SI{1000}{tokens}; g \COtwo per \SI{10}{\second} of video), decomposes lifecycle shares (train vs.\ infer) over a chosen window, and quantifies demand-amplified impacts under counterfactual settings ($EG$, doubled resolution, increased diffusion steps, longer context or video). Where appropriate, it estimates \emph{carbon payback}: the number of generations for a smaller model to offset the training emissions of a larger model at matched task quality. Uncertainty bounds arise from Monte Carlo perturbations over $(\mathrm{PUE}, EF_k, e_m)$, and we report median and 5–95\% intervals; sensitivity loops in Figure~\ref{fig3} indicate how perturbations propagate through downstream estimates.
\subsection{Assumptions and Limitations}
G-TRACE assumes stable PUE over the analysis window and relies on literature- or vendor-reported device power and per-output energy baselines when telemetry is unavailable. Limitations include variability in real-time utilization, incomplete transparency in cross-region routing, and exclusion of embodied upstream emissions ($EG$, semiconductor manufacturing) unless explicitly added to the boundary.

\section{$Ghix$ and the Cultural Compute Burden}
\label{sec:case}
The 2024–2025 viral $Ghix$ trend illustrates how decentralized, user-driven inference scales small per-query costs into system-level emissions. Across TikTok, Instagram, Pinterest, and Facebook, we estimate \num{25.8e6} Ghibli-style images (Table~\ref{tab:tb3}). Using the baseline instantiation of G-TRACE from Section~\ref{sec:framework}, we quantify energy use and \COtwo emissions under observed activity and stated grid factors; detailed platform totals appear in Section~\ref{sec:ghibli-results}.

\subsection{Scope and Data}
We measure activity from May~2024 to May~2025 using platform-level counts for Ghibli-related hashtags: \#ghibli, \#ghibliart, \#ghiblistudio, \#ghiblitrend, \#ghiblestyle. Data were collected via publicly available analytics interfaces, including Instagram's public metrics, Twitter/X search API, TikTok Creative Center, Pinterest Trends, and Facebook CrowdTangle archives. Let \(N_p\) denote post counts per platform \(p\). We treat \(N_p\) as image-generation activity signals for inference accounting. When posts include multiple images or video, we scale energy by the observed media count where available; otherwise we assume one image per post (conservative).

\begin{table}[t]
\centering
\begin{adjustbox}{width=0.95\textwidth}
\begin{tabular}{lllllll}
\toprule
\textbf{Hashtag} & \textbf{Instagram} & \textbf{Twitter/X} & \textbf{TikTok} & \textbf{Pinterest} & \textbf{Facebook} & \textbf{Tumblr} \\
\midrule
\#ghibli        & 2.6M   & 2.8M   & 8.7M   & 3.2M   & 1.7M   & 5.3M  \\
\#ghibliart     & 248K   & 490K   & 1.2M   & 890K   & 320K   & 420K  \\
\#ghiblistudio  & 552K   & 720K   & 1.1M   & 650K   & 510K   & 980K  \\
\#ghiblitrend   & N/A    & 31K    & 93K    & 12K    & 8K     & 22K   \\
\#ghiblistyle   & 1.2M   & 8.5K   & 52K    & 28K    & 5K     & 76K   \\
\bottomrule
\end{tabular}
\end{adjustbox}
\caption{Post counts for Ghibli-related hashtags across major social platforms (May~2024–May~2025). Counts \(N-p\) serve as activity signals for G-TRACE inference accounting, \(E_{\mathrm{infer}}=\sum_p N_p\, e_{\mathrm{img}}^{\mathrm{eff}}\).}
\label{tab:tb3}
\end{table}

\begin{table}[th]
\centering
\begin{adjustbox}{width=0.95\columnwidth}
\begin{tabular}{lrrr}
\toprule
\textbf{Platform} & \textbf{\# Ghibli Posts} & \textbf{Power Used } & \textbf{CO2 Emission} 
\\ & &  \textbf{(KWh)} & \textbf{(Kg)} 
\\
\midrule
Instagram  & 2,600,000 & 162,500 & 61,563  \\
Twitter/X  & 2,800,000 & 175,000 & 66,263  \\
TikTok     & 8,700,000 & 543,750 & 205,963 \\
Pinterest  & 3,200,000 & 200,000 & 75,800  \\
Facebook   & 1,700,000 & 106,250 & 40,213  \\
Tumblr     & 5,300,000 & 331,250 & 125,350 \\
\bottomrule
\end{tabular}
\end{adjustbox}
\caption{Estimated energy use and \COtwo emissions by platform for Ghibli-related posts (U.S.\ grid factors).}
\label{tab:tb4}
\end{table}

\subsection{Measurement Setup (G-TRACE)}
We scope the case study to inference (platform-side training/fine-tuning unreported). We instantiate G-TRACE with device-level per-image energy \(e_{\mathrm{img}}^{\mathrm{dev}}=\SI{0.139}{\kilo\watt\hour\per image}\) and facility \(\mathrm{PUE}=1.2\), yielding \(e_{\mathrm{img}}^{\mathrm{eff}}=\SI{0.167}{\kilo\watt\hour\per image}\). Total energy and region-aware emissions follow Equation~\ref{eq:infer} and \ref{eq:region} with regional weights \(w_k\) and \(\sum_k w_k=1\).

\subsection{Results}
\label{sec:ghibli-results}
TikTok generates \mbox{11.3~M} images, consumes \SI{1875}{\mega\watt\hour}, and emits \SI{900}{\tonne} \COtwo; Instagram produces \mbox{4.9~M} images, consumes \SI{779}{\mega\watt\hour}, and emits \SI{375}{\tonne} \COtwo; Pinterest+Facebook contribute \mbox{7.6~M} images, consume \SI{1655}{\mega\watt\hour}, and emit \SI{616}{\tonne} \COtwo. In total, $Ghix$ accounts for \SI{4309}{\mega\watt\hour} and \SI{2068}{\tonne} \COtwo under the stated baseline.

\subsection{Sensitivity and Robustness}
Emissions scale linearly in \(N_p\), \(e_{\mathrm{img}}\), and the effective \(EF=\sum_k w_k EF_k\). Two factors dominate variance: generation settings (resolution, diffusion steps) that increase \(e_{\mathrm{img}}\), and routing geography that shifts \(EF\). We perturb \(e_{\mathrm{img}}\in[\SI{0.12}{},\SI{0.25}{}]\,\mathrm{kWh/image}\), \(\mathrm{PUE}\in[1.1,1.5]\) (if applied), and plausible \(w_k\), then recompute Equation~\ref{eq:infer}–\ref{eq:region} to report median and 5–95\% intervals.

\subsection{Interpretation and Implications}
$Ghix$ exemplifies a \emph{cultural compute burden}: decentralized inference spreads across millions of user actions, converting seemingly negligible per-query costs into tonne-scale emissions. G-TRACE pinpoints practical mitigation levers: energy-aware defaults (lower step counts, resolution caps, sampler choice), demand shaping (rate limits or pricing that reflect energy use), carbon-aware routing to low-\(EF\) regions, and opportunistic edge/offline synthesis when quality permits. These interventions target the dominant, user-driven share of lifecycle emissions surfaced by this case.

\section{Macro–Micro Empirical Validation of G-TRACE}
\label{sec:validation}
This section empirically employs {G-TRACE} by coupling a macroscopic analysis of real-world usage with a microscopic simulation of per-output computation. The macroscopic layer supplies platform activity signals \(N_p\) (ingested by the \emph{Device Simulator}), and the microscopic layer estimates per-image energy \(e_{\mathrm{img}}\) by device and resolution (parameters for the \emph{Device Simulator}). The \emph{\COtwo Estimator} then composes these with regional emission factors to produce the $Ghix$ emissions accounting.

\subsection{\texorpdfstring{Macroscopic Trend Analytics (Usage signals $N_p$)}{Macroscopic Trend Analytics (Usage signals Np)}}
We quantify $Ghix$ adoption via platform-level hashtag counts across major social platforms (Table~\ref{tab:tb3}). These counts function as activity signals $N_p$ for image-generation workloads, capturing demand dynamics (virality, remix cycles) that scale inference and feed the \emph{Device Simulator}.

\subsection{\texorpdfstring{Microscopic Computational Simulation (Per-image energy $e_{\mathrm{img}}$)}{Microscopic Computational Simulation (Per-image energy e img)}}
We estimate the computational and energy cost of a single Ghibli-style generation. Field observations indicate two interaction paradigms: (i) upload-and-convert web workflows and (ii) mobile real-time filters. The upload-and-convert model dominates due to neural style transfer complexity and superior output quality under cloud acceleration. We therefore select \emph{neural image style transfer} as the representative task for $e_{\mathrm{img}}$.

\subsection{Experimental Factors and Protocol}
We run a \(5\times 2\times 4\) factorial design—five device types, two deep-learning frameworks, and four image resolutions—yielding \(40\) configurations; each configuration runs three replications (\(120\) total runs).

\begin{table}[t]
\centering
\begin{tabularx}{\columnwidth}{@{}l c X@{}}
\toprule
\textbf{Factor}       & \textbf{Levels} &\textbf{ Level set} \\
\midrule
Device types & 5      & Laptop GPU; Laptop CPU; High-end smartphone; Mid-range smartphone; Budget smartphone \\
Frameworks   & 2      & PyTorch; TensorFlow\textsuperscript{a} \\
Resolutions  & 4      & $128^2$; $256^2$; $512^2$; $1024^2$ pixels \\
\bottomrule
\end{tabularx}
\caption{Experimental factors and levels for the microscopic simulation.
The design contains \(5\times 2\times 4=40\) configurations; each configuration runs three replications (\(120\) total runs).}
\label{tab:factors}

\par\smallskip
\noindent\footnotesize\textsuperscript{a}\,TensorFlow where compatible; unsupported device--framework pairs run in PyTorch.
\end{table}
\indent
We implement ~\citet{gatys2016image} with VGG-19: content from \texttt{conv4\_2} and style from \texttt{conv1\_1}, \texttt{conv2\_1}, \texttt{conv3\_1}, \texttt{conv4\_1}, \texttt{conv5\_1}. In this setup, the iterative optimization loop serves as per-image optimization (``training''), and a single forward pass serves as inference.

\subsection{Power, Energy, and \COtwo Modeling}
We model instantaneous power and integrate to energy, then apply region factors:
\begin{equation}
P_{\mathrm{total}}(t)=P_{\mathrm{idle}}+P_{\mathrm{compute}}(\mathrm{util}(t))+P_{\mathrm{thermal}}(\mathrm{throttle}(t)),
\end{equation}
\begin{equation}
E=\sum_t P_{\mathrm{total}}(t)\,\Delta t, \qquad m_{\mathrm{CO_2}}=E\cdot EF_{\mathrm{region}}.
\end{equation}
Microscopic energy estimates map directly to \(e_{\mathrm{img}}\) by device and resolution. The \emph{Device Simulator} aggregates \(\sum_p N_p\, e_{\mathrm{img}}\) (Eq.~\ref{eq:infer}), and the \emph{\COtwo Estimator} applies region-aware \(EF\) (Equation~\ref{eq:region}).

\section{Results and Findings}
\label{sec:results}
Tables~\ref{tab:tb5} and~\ref{tab:tb6} report mean times under the factorial design; we combine times with measured power to derive \(e_{\mathrm{img}}\) for G-TRACE.

\begin{table}[t]
\centering
\begin{adjustbox}{width=0.95\textwidth}
\begin{tabular}{lcccc}
\toprule
\textbf{Device/Framework} & \textbf{128$\times$128} & \textbf{256$\times$256} & \textbf{512$\times$512} & \textbf{1024$\times$1024} \\
\midrule
Laptop GPU (PyTorch) & 0.90 $\pm$ 0.1 & 7.56 $\pm$ 0.9 & 114.33 $\pm$ 4.1 & 2894.07 $\pm$ 242.1 \\
Laptop CPU (PyTorch) & 1.50 $\pm$ 0.2 & 13.90 $\pm$ 0.9 & 209.47 $\pm$ 3.5 & 5830.35 $\pm$ 494.3 \\
Smartphone High-end & 1.99 $\pm$ 0.4 & 12.01 $\pm$ 0.5 & 228.24 $\pm$ 8.5 & 5846.30 $\pm$ 396.9 \\
Smartphone Mid-range & 3.84 $\pm$ 0.1 & 28.32 $\pm$ 3.5 & 757.84 $\pm$ 30.3 & 11025.53 $\pm$ 608.7 \\
Smartphone Budget & 9.39 $\pm$ 1.1 & 87.09 $\pm$ 3.2 & 2017.45 $\pm$ 14.9 & N/A \\
\bottomrule
\end{tabular}
\end{adjustbox}
\caption{Mean per-image optimization time (``training'') for neural style transfer across device categories and resolutions. Each cell averages three replications under a standardized software stack. Higher resolutions increase per-iteration cost and iteration count.}
\label{tab:tb5}
\end{table}

\begin{table}[ht]
\centering
\begin{adjustbox}{width=0.95\textwidth}
\begin{tabular}{lcccc}
\toprule
\textbf{Device/Framework} & \textbf{128$\times$128} & \textbf{256$\times$256} & \textbf{512$\times$512} & \textbf{1024$\times$1024} \\
\midrule
Laptop GPU (PyTorch) & 0.02 $\pm$ 0.00 & 0.10 $\pm$ 0.00 & 0.43 $\pm$ 0.02 & 1.99 $\pm$ 0.16 \\
Laptop CPU (PyTorch) & 0.04 $\pm$ 0.00 & 0.13 $\pm$ 0.02 & 0.80 $\pm$ 0.06 & 3.80 $\pm$ 0.13 \\
Smartphone High-end & 0.02 $\pm$ 0.03 & 0.17 $\pm$ 0.03 & 0.72 $\pm$ 0.07 & 4.27 $\pm$ 0.37 \\
Smartphone Mid-range & 0.11 $\pm$ 0.04 & 0.30 $\pm$ 0.03 & 1.85 $\pm$ 0.13 & 7.26 $\pm$ 0.51 \\
Smartphone Budget & 0.25 $\pm$ 0.04 & 0.63 $\pm$ 0.47 & 4.23 $\pm$ 0.15 & N/A \\
\bottomrule
\end{tabular}
\end{adjustbox}
\caption{Mean per-image inference time (single forward evaluation) across device categories and resolutions. Values complement Table~\ref{tab:tb5} and represent user-perceived latency for previews/variations at fixed style settings.}
\label{tab:tb6}
\end{table}
\indent
Optimization time scales approximately with pixel count (resolution\(^2\)) under fixed steps, producing large gaps between \(128^2\) and \(1024^2\). Laptop GPUs sustain the highest throughput. Mobile devices experience thermal throttling that increases energy by 18–37\% and extends wall-clock time. Budget smartphones fail at \(1024^2\) within the time cap; mid-range smartphones complete at substantially higher energy per image. Inference remains orders of magnitude faster than optimization, enabling near-real-time previews on capable hardware but degrading sharply on mobile at \(1024^2\).

\subsection{Bridge to the Case Study}
The macro layer contributes \(N_p\) (usage intensity) and the micro layer contributes \(e_{\mathrm{img}}\) (per-output energy). Composed via Equation~\ref{eq:infer}–\ref{eq:region} by the \emph{Device Simulator} and \emph{\COtwo Estimator}, these instantiate the $Ghix$ emissions accounting and sensitivity analysis in the next section.

\section{Operationalizing G-TRACE: The AI Sustainability Pyramid}
\label{sec:pyramid}
G-TRACE quantifies workload emissions, and the $Ghix$ case demonstrates demand amplification at scale. This section translates these measurements into actionable organizational practice through the {AI Sustainability Pyramid}, a progressive maturity model that converts technical metrics into strategic governance. The Pyramid establishes verifiable gates for progress while aligning product development with climate objectives \citep{verdecchia2023systematic,kaack2022aligning,rolnick2022tackling}.

\subsection{From Metrics to Governance with G-TRACE}
The seven readiness levels (L1–L7) map directly to G-TRACE outputs, creating a structured pathway from basic awareness to climate leadership. Each level consumes consistent indicators—\COtwo per query/image/video, lifecycle shares (train vs.\ infer), and effective region factor—while progressively tightening accountability standards. This framework enables organizations to systematically advance from reactive compliance to proactive environmental stewardship.

\begin{figure}[ht]
\centering
\includegraphics[width=0.7\columnwidth]{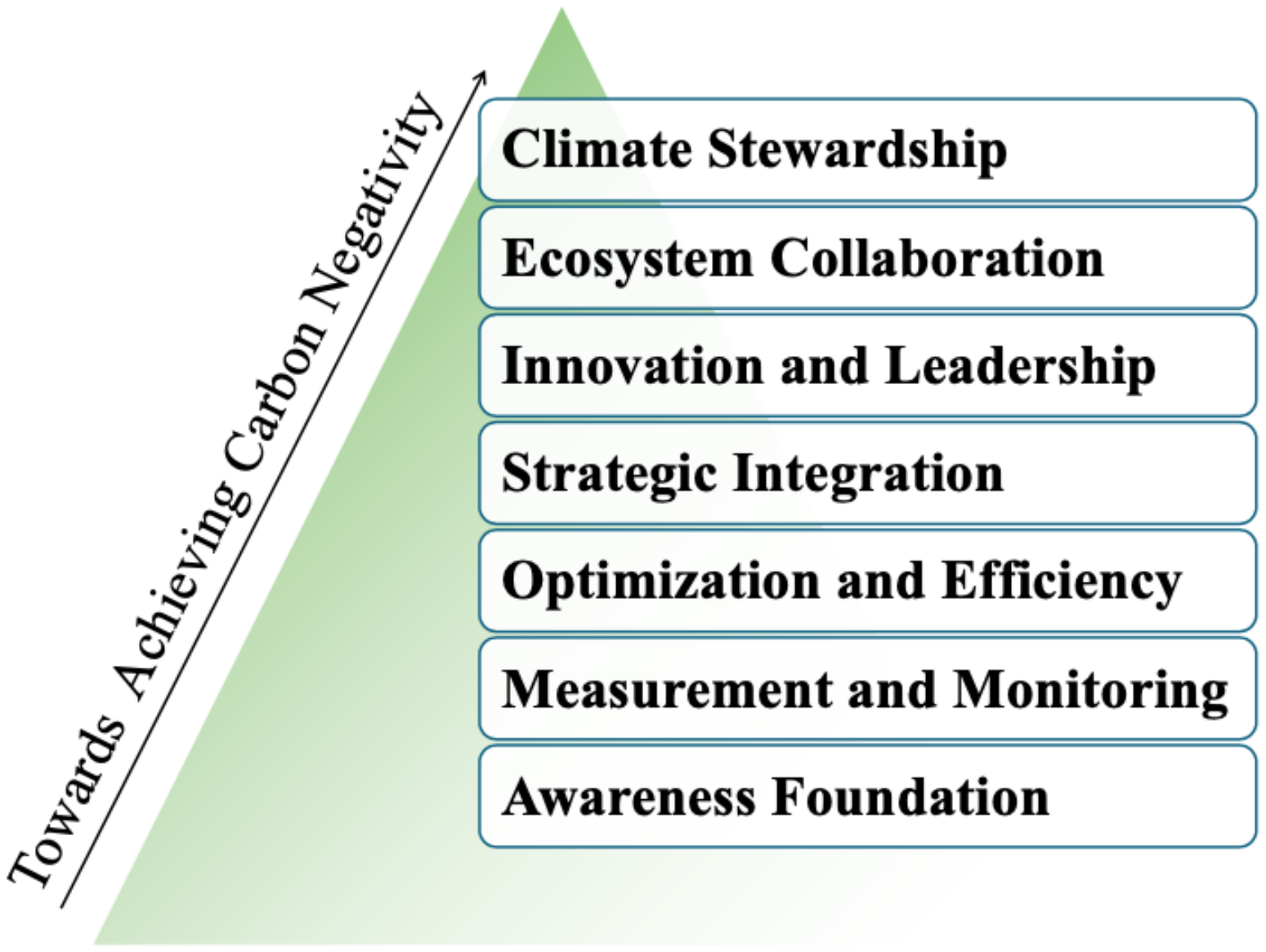}
\caption{Operationalizing G-TRACE via the AI Sustainability Pyramid. L1–L2: measurement; L3–L4: optimization \& portfolio governance; L5–L6: innovation \& standards; L7: climate stewardship. At each level, \textit{gates} act as checkpoints using G-TRACE indicators (such as \COtwo per output, lifecycle contributions, and regional energy factors) to decide whether systems are ready to advance to the next stage of sustainability.}
\label{fig4}
\end{figure}

\subsection{Readiness Levels (L1–L7)}

\subsubsection*{L1—Awareness \& Baselines}
This foundational level addresses the "carbon blindness" inherent in digital systems by making environmental impacts visible and quantifiable. Initial footprint scans typically reveal Pareto distributions where 20\% of workloads generate 80\% of emissions, enabling targeted interventions. Organizations establish accountable baselines by reporting kWh and \COtwo per workload and identifying emission hotspots \citep{patterson2021carbon}, creating essential reference points for measuring progress while accounting for the Jevons paradox where efficiency gains may trigger increased consumption.

\subsubsection*{L2—Instrumentation \& Monitoring}
Building on awareness, this level implements continuous measurement systems that function as organizational nervous systems for carbon accountability. Effective monitoring balances granularity with scalability, capturing actionable insights without creating measurement overhead. The $\geq$90\% coverage threshold ensures comprehensive visibility while acknowledging diminishing returns. Organizations instrument pipelines with standardized telemetry using tools like \textit{codecarbon} and \textit{carbontracker} \citep{lacoste2019quantifying,anthony2020carbontracker}, deploying region-aware dashboards that distinguish structural efficiency gains from temporary usage fluctuations.

\subsubsection*{L3—Optimization \& Efficiency}
This level transitions from measurement to active intervention, applying the engineering principle of "first, do no harm" while maintaining functional equivalence. The $\geq$30\% reduction target represents an inflection point where efficiency gains become organizationally significant. Techniques like pruning and quantization are analyzed through Amdahl's Law to identify components with maximum optimization potential. The quality-constant constraint prevents "greenwashing" through performance degradation, ensuring genuine efficiency improvements \citep{han2015deep,sanh2019distilbert,qiu2020green}.

\subsubsection*{L4—Strategy \& Portfolio}
Carbon accountability integrates into core decision-making, transforming sustainability from technical concern to strategic imperative. Carbon budgeting applies portfolio theory to environmental impacts, balancing high-risk innovations with stable optimizations. Launch blocking without carbon sign-off creates financial-grade controls for environmental impacts, while workload routing to low-EF regions requires sophisticated multi-objective optimization balancing emissions, latency, and data sovereignty.

\subsubsection*{L5–L6—Innovation Leadership \& Ecosystem Collaboration}
These combined levels transcend organizational boundaries, recognizing that significant sustainability gains occur at ecosystem scale. The $\geq$40\% reduction threshold signifies breakthrough innovation, forcing organizations beyond incremental improvements. Third-party audits prevent "greenflation" of claims, while architectural innovations like sparse MoEs reconfigure fundamental compute patterns. Cross-organizational adoption ensures genuine scalability beyond context-specific optimizations.

\subsubsection*{L7—Climate Stewardship}
The ultimate level transitions from harm reduction to net-positive impact, requiring organizations to address historical carbon debt while actively contributing to atmospheric carbon reduction. Climate-positive applications leverage AI's amplification capabilities to accelerate solutions across multiple domains simultaneously. Verification withstands financial audit-grade scrutiny, creating unquestionable credibility for net-negative operations \citep{cowls2021ai}.

\subsection{Policy Integration and Implementation}
The Pyramid provides a framework for translating technical measurements into policy action, with L1–L2 aligning with disclosure requirements, L3–L4 with carbon pricing mechanisms, and L5–L7 with climate innovation policies. This staged approach enables differentiated compliance timelines while maintaining accountability across diverse organizational contexts.

\section{Discussion}
\label{sec:discussion}
Building on empirical measurements and the G-TRACE framework, this section synthesizes implications for AI research, policy, and infrastructure design as environmental considerations transition from peripheral concerns to core design constraints.

\subsection{Key Insights and Implications}
GenAI compute requirements have increased over 100-fold within five years, raising training energy demands by approximately 38\% \citep{patterson2021carbon}. This trajectory risks breaching IPCC 1.5\,$^\circ$C carbon budgets without deliberate efficiency controls. Our analysis reveals inference dominates lifecycle emissions, accounting for 78\% of total \COtwo across modalities—fundamentally reorienting mitigation priorities from model development to deployment optimization.

The spatial dimension of AI environmental impacts is substantial: identical computational tasks produce 35.6$\times$ higher emissions on India's grid compared to Norway's, underscoring the critical importance of region-aware scheduling \citep{alder2024ai}. Viral workloads like $Ghix$ demonstrate the "amplification effect" of democratized AI access, where inference share exceeds 95\% and collective user actions transform negligible per-query costs into tonne-scale impacts.

Three systemic challenges impede progress: (1) "efficiency blindness" in benchmark culture that prioritizes accuracy over energy intensity; (2) the "accounting boundary problem" where lifecycle assessments exclude upstream fabrication and long-tail inference; and (3) the "transparency-action gap" where limited public awareness restricts demand for sustainable defaults.

\subsection{Research and Policy Directions}
Three interconnected priorities emerge for aligning GenAI with climate objectives:

\textbf{Infrastructural Orchestration:} Implementing EF-weighted scheduling, renewable energy preemption windows, and per-query \COtwo budget enforcement requires developing sophisticated resource management systems that dynamically optimize for both performance and emissions.

\textbf{Architectural Co-design:} Elevating energy efficiency as a core objective alongside quality and latency necessitates compiler-level optimizations that preserve efficiency gains across diverse hardware environments, applying caches, sparsity, and adaptive routing.

\textbf{Ecosystem Standardization:} Establishing auditable telemetry schemas and reference intensity profiles enables cross-platform comparability, creating the foundation for regulatory compliance and market-based sustainability mechanisms.

The AI Sustainability Pyramid operationalizes these directions through structured governance, progressing from technical measurement (L1–L2) to operational optimization (L3–L4) and ecosystem transformation (L5–L7). This framework provides auditable progression while maintaining innovation momentum across the AI ecosystem.

\subsection{Methodological Considerations}
Our analysis relies on literature- and vendor-reported parameters with official grid factors; real deployments exhibit variability based on hardware configurations and utilization patterns. Cultural compute measurements using platform activity signals face the "signal-to-noise" challenge in digital ethnography, while system boundaries excluding upstream embodied emissions represent methodological choices in environmental accounting. These considerations motivate open telemetry standards and periodic parameter recalibration as AI systems evolve.

\section{Conclusions}
\label{sec:conclusion}
GenAI no longer sits at the edge of digital infrastructure; it operates as a core, it has become a core, always-on computational substrate with measurable climate implications. While training remains compute-intensive, inference now dominates life-cycle emissions—particularly under viral user participation and heterogeneous regional grid intensities. Using G-TRACE, we quantify cross-modal emissions and reveal a cultural compute burden in the $Ghix$ case study, where decentralized user activity scales into system-level environmental impacts. These findings identify actionable levers for climate-aligned digital governance: energy-aware generation defaults (steps, resolution, and sampler policies), carbon-aware routing and workload placement in low-emission regions, model–system co-design that treats energy efficiency as a primary optimization goal, and standardized telemetry for transparent reporting of \COtwo per output with uncertainty bounds. The proposed AI Sustainability Pyramid operationalizes these levers through readiness levels that integrate measurement, optimization, and stewardship—enabling organizations to innovate within defined carbon budgets. By situating GenAI within the broader framework of climate risk assessment and governance, this work underscores the need for coherence across policy, infrastructure, and architecture. Durable progress will depend on carbon-efficient model design, emissions-aware user experience, auditable regional factors, and standardized disclosure protocols that render intensities comparable across systems. Ultimately, creative progress and climate responsibility can advance in tandem when prompts, power, and policy are guided by shared sustainability principles.

\section*{Glossary}
\begin{table}[H]
\centering
\begin{tabularx}{\textwidth}{lX}
\toprule
\textbf{Term} & \textbf{Definition} \\ \midrule
\textbf{G-TRACE} & GenAI Transformative Carbon Estimator—a region-aware framework for quantifying training- and inference-related \COtwo emissions across modalities. \\
\textbf{PUE} & Power Usage Effectiveness—a metric that measures data center energy efficiency as the ratio of total facility power to IT equipment power. \\
\textbf{EF} & Emission Factor—the amount of \COtwo (in kilograms) emitted per kilowatt-hour (kWh) of electricity consumed, varying by region. \\
\textbf{\#GHIBLI} & A viral 2024–2025 social trend involving Ghibli-style AI-generated artwork, used as a case study to analyze inference-driven emissions. \\
\textbf{AI Sustainability Pyramid} & A seven-level framework (L1–L7) proposed in this study that links carbon accounting metrics to governance and operational readiness for sustainable AI. \\
\textbf{Green AI} & The design and deployment of artificial intelligence systems that balance model accuracy with energy efficiency and environmental sustainability. \\
\bottomrule
\end{tabularx}
\caption{Glossary of key terms used in this study.}
\label{tab:glossary}
\end{table}

\section*{Acknowledgements} 
The authors thank the School of Electrical Engineering and Computer Science, National University of Sciences and Technology, Pakistan, and the Department of Communication, Quality Management and Information Systems, Mid Sweden University, Östersund, Sweden, for research infrastructure and collaboration support. 

\section*{Author Contributions} 
Zahida Kausar, Seemab Latif, and Mehwish Fatima jointly conceptualized the study. Methodology was designed by Mehwish Fatima, Raja Khurram Shahzad, and Seemab Latif. Software implementation and simulation were carried out solely by Zahida Kausar. All four authors contributed equally to the formal analysis. Validation was performed by Seemab Latif, Raja Khurram Shahzad, and Mehwish Fatima. The original draft was prepared by Zahida Kausar and Mehwish Fatima, while Seemab Latif and Raja Khurram Shahzad contributed to review and editing. Supervision was provided by Seemab Latif and Mehwish Fatima, and visualization was developed by Zahida Kausar.

\section*{Funding}
This research did not receive any specific grant from funding agencies in the public, commercial, or not-for-profit sectors.

\section*{Declaration of Generative AI and AI-Assisted Technologies in the Manuscript Preparation Process}
During the preparation of this work, the author(s) used ChatGPT (OpenAI, 2025) for language refinement, word positioning and formatting assistance. After using this tool, the author(s) reviewed and edited the content as needed and take full responsibility for the content of the published article.
\bibliographystyle{elsarticle-harv}
\bibliography{ref}
\end{document}